\documentclass[12pt,dvips]{article}

\usepackage{rotating}
\usepackage{epsfig}
\usepackage{color}

\newcommand{\wt}{\widetilde}
\newcommand{\imag}{\Im {\rm m}}

\newcommand{\mto}{m^2_{\tilde{t}_1}}
\newcommand{\mttt}{m^2_{\tilde{t}_2}}
\newcommand{\mbo}{m^2_{\tilde{b}_1}}
\newcommand{\mbt}{m^2_{\tilde{b}_2}}
\newcommand{\ghat}{\hat{g}^2}
\newcommand{\shat}{\hat{s}}

\newcommand{\lessim}{\stackrel{<}{{}_\sim}}
\newcommand{\gtrsim}{\stackrel{>}{{}_\sim}}

\begin{document}
\renewcommand{\thefootnote}{\fnsymbol{footnote}}

\mbox{ } \\[-1cm]
\mbox{ }\hfill KEK--TH--785\\
\mbox{ }\hfill hep--ph/0110138\\
\mbox{ }\hfill \today\\

\bigskip
\bigskip
\bigskip
\bigskip
\bigskip

\begin{center}
{\Large{\bf Observability of the Lightest MSSM Higgs Boson\\[2mm]
            with Explicit CP Violation via Gluon Fusion\\[2mm]
	    at the LHC}}\\[2cm]
            S.Y. Choi$^1$,\, Kaoru Hagiwara$^2$ and Jae Sik Lee$^2$ 
\end{center}

\bigskip

{\small
\begin{enumerate}
\item[{}] $^1$ Department of Physics, Chonbuk National University, Chonju
               561--756, Korea
\item[{}] $^2$ Theory Group, KEK, Tsukuba, Ibaraki 305--0801, Japan 
\end{enumerate}
}
\bigskip
\bigskip
\bigskip
\bigskip
\bigskip
\bigskip
\bigskip

\begin{abstract}
We investigate the observability of the lightest Higgs boson in the
gluon--fusion channel at the CERN Large 
Hadron Collider (LHC) in the minimal supersymmetric Standard Model with 
explicit CP--violating mixing among three neutral Higgs bosons. 
The lightest Higgs boson with its mass less than 130 GeV can be detected at
the LHC via its gluon--fusion production followed by the decay into two 
photons. The explicit CP violation can suppress both the production cross
section and  the two--photon
decay branching fraction so significantly that the signal cross section may be
more than ten times smaller than the SM signal. This reduction factor can be
as small as $1/40$ if the lightest Higgs boson mass is 115 GeV and its
production cross section at LEP2 is more than 90 \% that of the SM case.
\end{abstract}

\vskip 0.4cm



\newpage

The soft CP violating Yukawa interactions in the minimal supersymmetric 
standard model (MSSM) cause the CP--even and CP--odd neutral Higgs bosons 
to mix through loop corrections \cite{AP,CDL}. The loop--induced 
CP violation in the MSSM Higgs sector can be large enough to affect Higgs 
phenomenology at present and future colliders significantly 
\cite{AP,EXCP_FC,CL1,CL11,CLPLC,CL4}. 
In this letter, we study the effects of the 
CP--violating mixing on the production of the lightest neutral MSSM Higgs boson 
through gluon fusion and its decay into a photon pair,
which is of crucial importance for detecting the lightest Higgs
boson with its mass less than 130 GeV at the LHC \cite{Orsay}.
After briefly reviewing the loop--induced CP--violating 
mixing \cite{CDL} of three neutral Higgs bosons in the MSSM,  
we estimate the effects of the CP phases on the production of the 
lightest Higgs boson though gluon fusion \cite{Dedes}
and its branching fraction of the two--photon decay mode, respectively. 
Finally, combining both the
production and decay of the lightest Higgs boson, we discuss the
observability of the lightest MSSM Higgs boson at the LHC in
the presence of explicit CP violation in the Higgs sector.

The loop--induced CP--violating neutral Higgs boson mixing is determined
by the Higgs boson mass matrix obtained by taking all the second derivatives 
of the effective MSSM Higgs potential \cite{CW,OKADA}
\begin{eqnarray}
\label{e2}
V_{\rm Higgs}\hskip -0.3cm 
   &= \, \frac{1}{2}m_1^2\left(\phi_1^2+a_1^2\right)
     +\frac{1}{2}m_2^2\left(\phi_2^2+a_2^2\right)
     -\left|m_{12}^2\right|\left(\phi_1\phi_2-a_1 a_2\right) 
           \cos (\xi + \theta_{12}) \nonumber \\ 
   & +\left|m^2_{12}\right|
      \left(\phi_1 a_2 +\phi_2 a_1\right)\sin(\xi+\theta_{12})
     +\frac{\ghat}{8} {\cal D}^2 
     +\frac{1}{64\,\pi^2} {\rm Str} \left[
           {\cal M}^4 \left(\log\frac{{\cal M}^2}{Q^2} 
	                  - \frac{3}{2}\right)\right] \,,
\end{eqnarray}
where ${\cal D} = \phi_2^2 + a_2^2 - \phi_1^2 - a_1^2$,
$\ghat = (g^2+g'^2)/4$ with the SU(2)$_L$ and U(1)$_Y$
gauge couplings $g$ and $g'$, and $\phi_i$ and $a_i$ 
($i=1,2$) are the neutral components of the two 
Higgs doublet fields:
\begin{eqnarray}
\label{e1}
H_1^0 = \frac{1} {\sqrt{2}} \left( \phi_1 + i a_1 \right)\,, \ \ \ \ \
H_2^0 = \frac {{\rm e}^{i \xi}} {\sqrt{2}} \left( \phi_2 + i a_2 \right)\,.
\end{eqnarray}
All the tree--level parameters of the Higgs potential (\ref{e2}) such as 
$m_1^2, \ m_2^2 $ and $m_{12}^2=\left|m^2_{12}\right|{\rm e}^{i\theta_{12}}$ 
are the running parameters evaluated at the renormalization scale $Q$, 
rendering the Higgs potential (almost) independent of $Q$ up to 
two--loop--order corrections. The super--trace is to be taken over all the 
bosons and fermions that couple to the Higgs fields.

The matrix ${\cal M}$ in Eq.~(\ref{e2}) is the field--dependent mass matrix 
of all modes that couple to the Higgs bosons. The dominant contributions 
in the MSSM come from the third generation quarks and squarks because of their 
large Yukawa couplings. The field--dependent masses of the bottom and
top quarks are given by $m_b^2 = |h_b|^2 |H^0_1|^2$ and 
$m_t^2=|h_t|^2 |H^0_2|^2$ with the bottom and top Yukawa couplings $h_b$ 
and $h_t$, and the bottom-- and top--squark mass matrices read
\begin{eqnarray}
{\cal M}_{\tilde t}^2 &= \mbox{$ \left( \begin{array}{cc} 
m^2_{\wt Q} + m_t^2 - \frac{1}{8} \left( g^2 - \frac{g'^2}{3} \right) {\cal D}
\,\,\,\,\,\,\,\,\,\,
&
- h_t^* \left[ A_t^* \left(H_2^0 \right)^* + \mu H_1^0 \right] \\
- h_t \left[ A_t H^0_2 + \mu^* \left( H_1^0 \right)^* \right] &
m^2_{\wt U} + m_t^2 - \frac{g'^2}{6} {\cal D}
\end{array} \right)\,, $} \nonumber\\
\nonumber \\ 
{\cal M}_{\tilde b}^2 &= \mbox{$ \left( \begin{array}{cc} 
m^2_{\wt Q} + m_b^2 + \frac{1}{8} \left( g^2 + \frac{g'^2}{3} \right) {\cal D}
\,\,\,\,\,\,\,\,\,\,
&
- h_b^* \left[ A_b^*  \left( H_1^0 \right)^* + \mu H_2^0 \right] \\
- h_b \left[ A_b H_1^0 + \mu^* \left( H_2^0 \right)^* \right] &
m^2_{\wt D} + m_b^2 + \frac{g'^2}{12} {\cal D}
\end{array} \right)\,. $} 
\label{e4}
\end{eqnarray}
where $m^2_{\wt Q}, \ m^2_{\wt U}$ and
$m^2_{\wt D}$ are the real soft SUSY--breaking squark mass
parameters, $A_b$ and $A_t$ are the
complex soft SUSY--breaking trilinear parameters, and $\mu$ is the complex
supersymmetric Higgsino mass parameter.

The second derivatives of the potential, giving the mass matrix of the Higgs 
bosons (at vanishing external momenta), are then evaluated at its minimum 
point $\left(\phi_1,\, \phi_2,\, a_1,\, a_2\right)
=\left(v\cos\beta,\, v\sin\beta,\, 0,\, 0\right)$
with $v\simeq 246 \ {\rm GeV}$
and $\tan\beta = \langle \phi_2\rangle/\langle \phi_1\rangle$. 
After absorbing a Goldstone mode $G^0=a_1\cos\beta-a_2\sin\beta$ 
into the $Z$ boson, we are left with a real and symmetric 3$\times$3 
mass--squared matrix ${\cal M}_H^2$ of three physical states, 
$a \,(= a_1 \sin \beta + a_2 \cos \beta), \ \phi_1$ and $\phi_2$.
The two CP--violating entries of the symmetric matrix, which mix $a$ 
with $\phi_1$ and $\phi_2$, are given by
\begin{eqnarray}
\label{e13}
\left. {\cal M}^2_H \right|_{a \phi_1}
   = \frac{3}{16 \pi^2} \left\{
       \frac{m_t^2\Delta_{\tilde t}}{\sin \beta}\, F_t 
      +\frac{m_b^2\Delta_{\tilde b}}{\cos \beta}\, G_b\right\}, 
       \quad 
\left. {\cal M}^2_H \right|_{a \phi_2}
   = \frac{3}{16 \pi^2} \left\{
       \frac{m_t^2 \Delta_{\tilde t}}{\sin \beta}\, G_t
      +\frac{m_b^2 \Delta_{\tilde b}}{\cos \beta}\, F_b\right\}. 
\end{eqnarray}
The explicit forms of the dimensionless quantities $F_{t,b}$ and $G_{t,b}$ 
and all the CP--preserving entries of the
mass--squared matrix ${\cal M}^2_H$ can be found in Ref.~\cite{CDL}.
The rephasing--invariants
\begin{eqnarray}
\label{e9}
\Delta_{\tilde t}=\frac{\imag(A_t \mu {\rm e}^{i \xi})}
                  {\mttt - \mto} \,, \ \ \qquad
\Delta_{\tilde b}=\frac{\imag(A_b \mu {\rm e}^{i \xi})}
                  {\mbt - \mbo} ,
\end{eqnarray}
measure the amount of CP violation in the top and bottom squark--mass
matrices and vanish in the CP--invariant theories, leading to 
$|m^2_{12}|\,\sin(\xi+\theta_{12})=0$ in the potential (\ref{e2}). 
The matrix ${\cal M}^2_H$ can be 
diagonalized by an orthogonal matrix $O$;
\begin{eqnarray}
\label{OMIX}
O^T{\cal M}^2_H\, O={\sf diag}\,(m^2_{H_1},m^2_{H_2},m^2_{H_3}),
\end{eqnarray}
%
where the three mass-eigenvalues are ordered as
$m_{H_1}< m_{H_2}< m_{H_3}$. 

The loop--corrected neutral--Higgs--boson sector depends on
many parameters in the Higgs and squark sectors; a 
loop--corrected pseudoscalar mass $m_A$, $\tan\beta$, 
$\mu$, $A_t$, $A_b$, the scale $Q$, and the soft--breaking masses, 
$m_{\tilde Q}$, $m_{\tilde U}$, and $m_{\tilde D}$, as well as on 
the complex gluino--mass parameter $M_{\wt g}$ through one--loop corrections 
to the top and bottom quark masses \cite{SUSYHBB}. However, the CP violation 
in the Higgs sector is determined essentially by the rephasing 
invariant combinations $A_t \mu {\rm e}^{i\xi}$ and $A_b \mu {\rm e}^{i\xi}$, 
see Eq. (\ref{e9}), and is 
dominantly by the top--squark sector if $\tan\beta\leq 10$.
Therefore, we take in our numerical analysis the following parameter set:
\begin{eqnarray}
&& |A_t|=|A_b| = \kappa\,M_{\rm SUSY}\,, 
   \hspace{3.2 cm} |\mu|=2\,|A_t|\,,\nonumber \\
&& m_{{\widetilde Q},{\widetilde U},{\widetilde D}}=|M_{\widetilde g}|
   =M_{\rm SUSY} = 0.5~{\rm TeV}\,, \qquad {\rm Arg}(M_{\widetilde g})=0\,,
   \nonumber \\
&& \Phi \equiv {\rm Arg}(A_t\mu {\rm e}^{i\xi}) = 
{\rm Arg}(A_b\mu {\rm e}^{i\xi})\,. 
\label{eq:PARA}
\end{eqnarray}
%
Then, we vary the dimensionless parameter $\kappa$, the common
phase $\Phi$ and $\tan\beta$ in the numerical analysis, for which
the pseudoscalar mass parameter $m_A$ is chosen to fix the
the lightest Higgs boson mass $m_{H_1}$. Clearly, a large $\kappa$ implying 
large values of $|A_{t,b}|$ leads to  large CP--violating effects as clearly 
seen from Eq.~(\ref{e9}). However, $\kappa$ cannot be too large, because
it generates an unacceptably large value of the electron and neutron 
electric dipole moments (EDM's)\footnote{It is possible that the stringent 
two--loop EDM constraints may be satisfied by a cancellation among 
various contributions \cite{CEPW2,P-EDM,CL11}.} at the two--loop level
through the one--loop 
effective CP--odd couplings of the Higgs boson to the gauge bosons \cite{CKP}. 
Moreover, in order to avoid a color and electric--charge 
breaking minimum deeper than the electroweak vacuum, $\kappa$ cannot be 
significantly larger than the unity \cite{CCB-OLD,CLM}.

In the presence of the CP--violating neutral Higgs--boson mixing,
the amplitude for the resonance production $gg\rightarrow H_i$ ($i=1,2,3$) 
can be written as
\begin{eqnarray}
{\cal M}_{ggH_i}=\frac{m_{H_i}\alpha_s \delta_{ab}}{4\pi}
       \left\{S^g_i(m_{H_i})\left(\epsilon_1\cdot\epsilon_2
             -\frac{2k_1\cdot\epsilon_2\,k_2\cdot\epsilon_1}{m_{H_i}^2}\right) 
             -P^g_i(m_{H_i})\frac{2}{m_{H_i}^2}
\epsilon_{\mu\nu\rho\sigma}\,\epsilon_1^\mu \epsilon_2^\nu k_1^\rho
k_2^\sigma\right\},
\end{eqnarray}
where $a,b=(1\,\,{\rm to}\,\,8)$ are the color indices for the 
eight gluon fields, and $k_{1,2}$ and $\epsilon_{1,2}$ 
are the momenta and polarization vectors of two colliding gluons, respectively.
The scalar and pseudo--scalar
form factors are then given by
\begin{eqnarray}
&& S^g_i(m_{H_i})=\sum_f \Bigg\{g_{sf}^i\frac{m_{H_i}}{m_f} F_{sf}(\tau_{if})
                +\frac{1}{4}\sum_{j=1,2}g_{\tilde{f}_j\tilde{f}_j}^i
                 \frac{m_{H_i}}{m_{\tilde{f}_j}^2} 
		  F_0(\tau_{i\tilde{f}_j}) \Bigg\}\,, \nonumber \\
&& P^g_i(m_{H_i})=\sum_f g_{pf}^i\frac{m_{H_i}}{m_f} F_{pf}(\tau_{if}) \,,
\end{eqnarray}
where $\tau_{ix}=m_{H_i}^2/4m_x^2$, $g_{sf}^i$ and $g_{pf}^i$ are the couplings 
of the Higgs boson $H_i$ to the scalar and pseudo--scalar fermion bilinears
$\bar{f}f$ and $i\,\bar{f}\gamma_5 f$, respectively.
The CP--violating Higgs mixing leads to a simultaneous existence
of these two couplings. On the other hand, $g_{\tilde{f}_j\tilde{f}_j}^i$ 
is the coupling of $H_i$ to a diagonal sfermion pair. 
We refer to Ref.~\cite{CLPLC} for the explicit forms of
the couplings as well as the form factors $F_{sf}$, $F_{pf}$, and $F_0$.
Note that in the minimal SM only the scalar 
form factor due to the top--quark and bottom--quark contributions survives.

\begin{center}
\begin{figure}[htb]
  \vspace*{-.5cm}
  \hspace*{3.0cm} 
  \epsfxsize=11cm \epsfbox{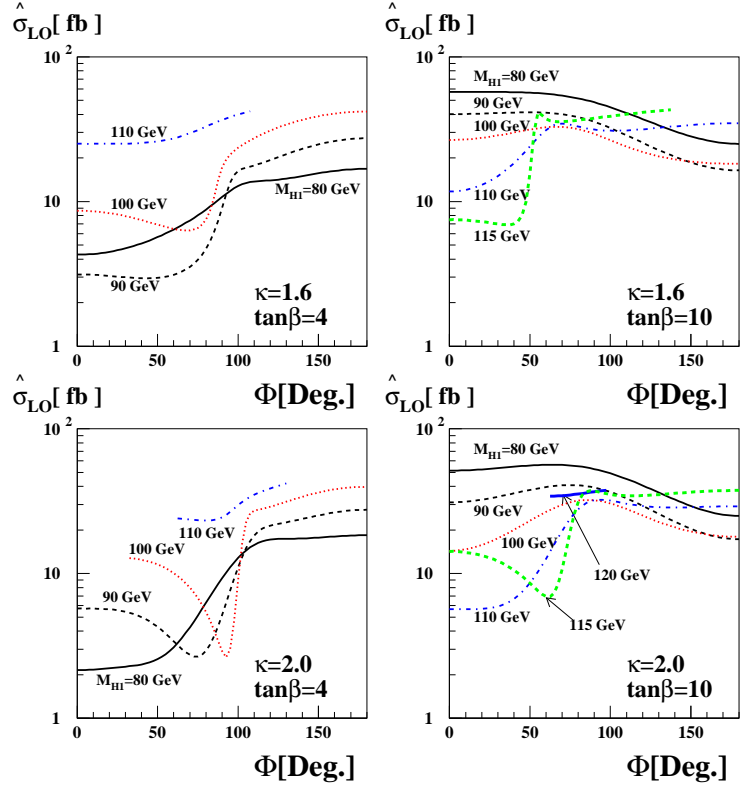}
  \vspace*{-.5cm}
\caption{\it The LO parton--level cross section of the lightest Higgs boson
             as a function of the phase $\Phi$ for 
             $m_{H_1}=80$ GeV (solid line), 90 GeV 
	     (dashed line), 100 GeV (dotted line), 110 GeV (dash--dotted
             line), 115 GeV (thick dashed line), and 120 GeV 
	     (thick solid line). We take the parameter set (\ref{eq:PARA}) 
	     with $\kappa=1.6$ (upper) and $\kappa=2.0$ (lower) and 
             two values of $\tan\beta=4$ (left) and 10 (right).}
\label{sigma0}
\end{figure}
\end{center}

The production cross section of a neutral Higgs boson
$H_i$ in $gg$ fusion is given by
\begin{eqnarray}
\sigma(gg\rightarrow H_i)=\frac{\alpha_s^2}{256\pi m_{H_i}^2}
      \left[\,\left|S^g_i\right|^2+\left|P^g_i\right|^2\right]\,
	   \delta(1-\frac{m_{H_i}^2}{\shat})\, 
\equiv \,\hat{\sigma}_{\rm LO}(gg\rightarrow H_i)
           \,\delta(1-\frac{m_{H_i}^2}{\shat}) \,,
\end{eqnarray}
with $\sqrt{\shat}$ the two--gluon c.m. energy. Figure~\ref{sigma0} shows 
the leading--order (LO) parton--level cross section 
$\hat{\sigma}_{\rm LO}(gg\rightarrow H_1)$ 
as a function of the phase $\Phi$ for $m_{H_1}=80$ GeV 
(solid line), 90 GeV (dashed line), 100 GeV (dotted line), 110 GeV 
(dash--dotted line), 115 GeV (thick dashed line), and 120 GeV 
(thick solid line) with the parameter set (\ref{eq:PARA}) for 
$\kappa=1.6$ (upper) and $2.0$ (lower) and for $\tan\beta=4$ (left) and 
10 (right), respectively. Note that the cases with $m_{H_1}\leq 100$ GeV
are also considered because the lightest Higgs boson $H_1$ could be
undetected at LEP2 with the $Z ZH_1$ coupling suppressed for 
non--vanishing $\Phi$.

The SM LO parton-level cross section at 
$m_{H_{\rm SM}}=m_{H_1}$ is $45.0~{\rm fb}\leq \hat{\sigma}^{\rm SM}_{\rm LO}
\leq 46.6~{\rm fb}$ for $m_{H_{\rm SM}}$ between 80 GeV and 120 GeV,
implying that the SM cross section does not depend on the Higgs boson 
mass significantly. On the contrary, the MSSM cross section is 
very sensitive to $\Phi$ and $m_{H_1}$ for both $\kappa=1.6$ and $2.0$ and 
for both $\tan\beta=4$ and 10. 
In particular, the cross section is significantly smaller than the SM 
one for small $\Phi$ and $\tan\beta$. This is due to the 
suppression of the coupling of the lightest MSSM Higgs boson to top 
quarks and to the cancellations between the fermionic and bosonic 
contributions. The cancellation is more significant when the
top--squark mass splitting is larger,
for smaller $\Phi$, larger $\kappa$ and smaller $\tan\beta$.
For $\kappa =1.6$, a significant cancellation between the fermionic 
and bosonic contributions occurs for all the  
Higgs mass cases at $\tan\beta=4$ when  $\Phi\lessim 70^{\rm o}$,
and for $m_{H_1}=115$ GeV (thick dashed line) at $\tan\beta=10$ 
when $\Phi\lessim 50^{\rm o}$.
For $\kappa=2.0$ and $\tan\beta=4$, a significant
cancellation occurs for $\Phi\lessim 90^{\rm o}$, while for $\kappa=2.0$ 
and $\tan\beta=10$ it occurs when $m_{H_1} = 110$ GeV and 115 GeV 
for $\Phi \lessim 70^{\rm o}$. 
In all cases, the cancellation between fermionic and bosonic contributions is
suppressed for $\Phi \gtrsim 120^{\rm o}$, reflecting the suppressed 
sfermion mass splitting. We find that
for $\tan\beta=4$ and $m_{H_1}\leq 100$ GeV the lightest Higgs boson has 
a large CP--odd component  when 
$70^{\rm o}\lessim \Phi \lessim 110^{\rm o} \,(\kappa=1.6)$ and 
$80^{\rm o}\lessim \Phi \lessim 130^{\rm o} \,(\kappa=2.0)$. 
Finally, since the scalar and pseudoscalar couplings of $H_1$ to the top quarks
are $g_{pt}^1\sim O_{11}/\tan\beta$ and $g_{st}^1\sim O_{31}/\sin\beta$,
respectively,
the top--quark loop contribution for the lightest Higgs boson of a large
CP--odd mixture is suppressed by $\cos\beta$ as compared to the case of
a pure CP--even lightest Higgs boson at $\Phi=0^{\rm o}/180^{\rm o}$.

\vskip -0.5cm

For a realistic estimate of the production cross section it is necessary
to include the next--to--leading--order (NLO) QCD loop
correction, denoted by the $\tan\beta$--dependent
$K$ factor to a good approximation \cite{DDS}; for small $\tan\beta$, it is 
1.5$-$1.7 and for large $\tan\beta$ it is in general close to 
unity except when the lightest Higgs boson approaches the SM limit, for which
$K\approx 1.5$.
In addition to the QCD NLO correction, we need to fold
the parton--level cross section with the gluon distribution function
to obtain the hadronic level cross section as 
\begin{eqnarray}
\sigma(pp\rightarrow H_1)=K\, \hat{\sigma}_{LO}(gg\rightarrow H_1)\,\, \tau
     \frac{{\rm d}{\cal L}^{gg}_{LO}}{{\rm d}\tau}\,,
\end{eqnarray}
where $\tau=m_{H_i}^2/s$ with $\sqrt{s}$ the hadron collider c.m. energy.
At the LHC with $\sqrt{s}=14$ TeV, the size of  the gluon fusion luminosity
factor ($\tau \frac{{\rm d}{\cal L}^{gg}_{LO}}{{\rm d}\tau}$) is 
between $0.6 \times 10^3$ and $0.3 \times 10^3$ 
for $m_{H_1}=80-130$ GeV \cite{CTEQ4m}.

In the presence of the radiatively induced CP--violating neutral Higgs 
boson mixing, the amplitude for the decay $H_i \rightarrow \gamma\gamma$ 
($i=1,2,3$) is written as
\begin{eqnarray} \label{hipp}
{\cal M}_{\gamma\gamma H_i}=\frac{m_{H_i}\alpha}{4\pi}
      \left\{S^\gamma_i(m_{H_i})
      \left(\epsilon^*_1\cdot\epsilon^*_2-\frac{2\, k_1\cdot\epsilon^*_2
            \,k_2\cdot\epsilon^*_1}{m_{H_i}^2}\right)
            -P^\gamma_i(m_{H_i})\frac{2}{m_{H_i}^2}
            \epsilon_{\mu\nu\rho\sigma}\,\epsilon_1^{*\mu}
	    \epsilon_2^{*\nu} k_1^\rho k_2^\sigma\right\}\,, 
\end{eqnarray}
where $k_{1,2}$ and $\epsilon_{1,2}$ are the momenta and polarization vectors
of the two photons, respectively. The scalar and pseudoscalar
form factors due to the (s)quark, $W^\pm$ and charged Higgs 
loops read
\begin{eqnarray}
S^\gamma_i(m_{H_i})&=&2N_C\sum_{f}Q_f^2\left\{
      g_{sf}^i\frac{m_{H_i}}{m_f} F_{sf}(\tau_{if})
     +\frac{1}{4}\sum_{j=1,2}g_{\tilde{f}_j\tilde{f}_j}^i
      \frac{m_{H_i}}{m_{\tilde{f}_j}^2} F_0(\tau_{i\tilde{f}_j})\right\}
      \nonumber \\
&&\hspace{0.5 cm}+
      \frac{g m_{H_i}}{2m_W}(c_\beta O_{2\,i}+s_\beta O_{3\,i})F_1(\tau_{iW})
     +\frac{v m_{H_i} C_i}{2 M_{H^\pm}^2} F_0(\tau_{iH^\pm})
      \,, \nonumber \\
P^\gamma_i(m_{H_i})&=&2N_C\sum_{f=t,b} Q_f^2g_{pf}^i\frac{m_{H_i}}{m_f}
      F_{pf}(\tau_{if})\,,
\label{eq:rrForm}
\end{eqnarray}
where $N_C=3$, $Q_f$ is the electric charge of the (s)fermion $f(\tilde{f})$
in the unit of the positron charge, and $C_i$ is the coupling of $H_i$ to 
the charged Higgs boson pair: ${\cal L}_{H_iH^+H^-}=v\,C_i\,H_iH^+H^-$.
We refer again to Ref.~\cite{CLPLC} for the explicit forms of the form factors
and the couplings $C_i$'s.  
The possible chargino contributions are  neglected by assuming that the
chargino states are very heavy. Note that the SM pseudoscalar form factor 
$P^\gamma_{\rm SM}$ vanishes and the SM scalar form factor $S^\gamma_{\rm SM}$ 
has only the top--quark and  and $W^\pm$--boson loop contributions.

\begin{center}
\begin{figure}[htb]
  \vspace*{-.5cm}
  \hspace*{3.0cm} 
  \epsfxsize=12cm \epsfbox{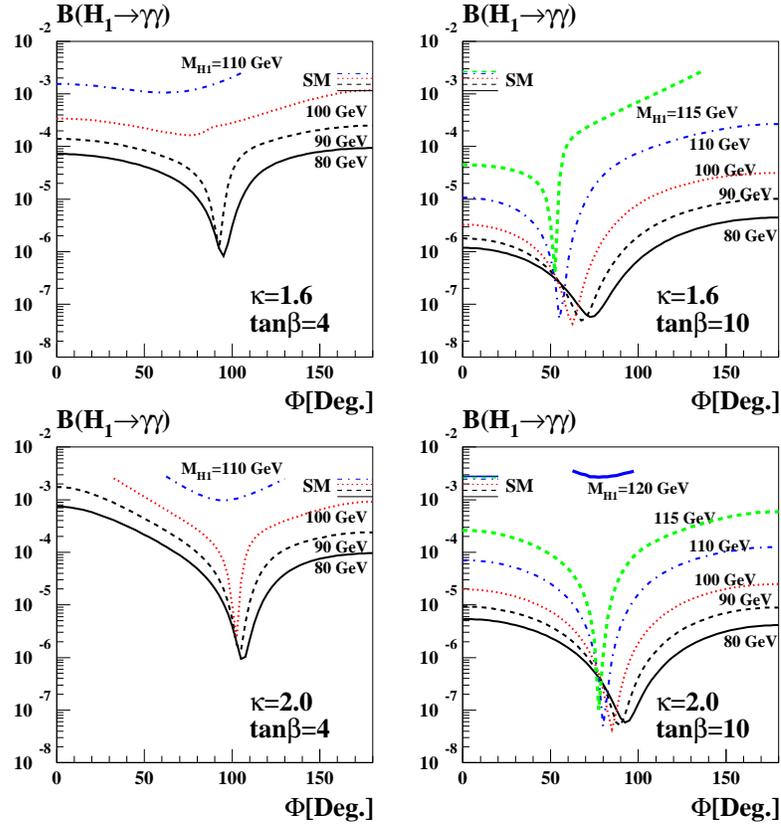}
  \vspace*{-.5cm}
\caption{\it The branching fraction ${\cal B}(H_1\rightarrow \gamma\gamma)$
         as a function of $\Phi$ for 
         $m_{H_1}=80$ GeV (solid line), 90 GeV (dashed line), 100 GeV 
	 (dotted line), 110 GeV (dash--dotted line), 115 GeV (thick dashed 
	 line), and 120 GeV (thick solid line). The parameter set 
	 (\ref{eq:PARA}) is with $\kappa=1.6$ (upper) and $\kappa=2.0$ (lower) 
	 and with $\tan\beta=4$ (left) and 10 (right).  
         The SM branching fraction is also shown
         with the same line convention at $m_{H_{\rm SM}}=m_{H_1}$. }
\label{br1}
\end{figure}
\end{center}

The decay width $\Gamma(H_i\rightarrow\gamma\gamma)$ is then given by
\begin{eqnarray}
\Gamma(H_i\rightarrow \gamma\gamma)=\frac{m_{H_i}\alpha^2}{256\pi^3}
         \left[\,\left|S^\gamma_i(m_{H_i})\right|^2
              +\left|P^\gamma_i(m_{H_i})\right|^2\right]\,,
\end{eqnarray}
in terms of the scalar and pseudoscalar form factors in Eq.~(\ref{eq:rrForm}). 
The main contribution to the decay of the lightest MSSM Higgs boson 
into two photons is from the $W^\pm$--boson loop giving rise to the
scalar form factor $S^\gamma_1(m_{H_1})$, which is determined
by the coupling of $H_1$ to $W^+W^-$.
We find that the $H_1W^+W^-$ coupling is very sensitive to the CP--violating 
neutral Higgs boson mixing. For example, if $H_1$ is a pure CP--odd state, 
the coupling vanishes. As a result, the partial decay width and its
branching fraction can be significantly suppressed in the presence of 
the CP--violating phases. Figure~\ref{br1} shows the the branching
fraction ${\cal B}(H_1\rightarrow \gamma\gamma)$ as a function of the 
phase $\Phi$ for $m_{H_1}=80$ GeV (solid
line), 90 GeV (dashed line), 100 GeV (dotted line), 110 GeV (dash--dotted
line), 115 GeV (thick dashed line), and 120 GeV (thick solid line) for
the parameter set (\ref{eq:PARA}) with $\kappa=1.6$ (upper) and 
$\kappa=2.0$ (lower) and with $\tan\beta=4$ (left) and 10 (right). 
For reference, the SM branching fraction, which turns out to be 
between $1\times 10^{-3}$ and $3\times 10^{-3}$,
is also shown in each frame with the same line convention, fixing 
$m_{H_{\rm SM}}=m_{H_1}$. The MSSM branching fraction is so
sensitive to the phase $\Phi$ that it is suppressed by a factor of $10^3$ 
around $\Phi= 100^{\rm o}\, (\tan\beta=4)$ and $10^4$ around 
$\Phi= 50^{\rm o}-75^{\rm o} \, (\tan\beta=10, \kappa=1.6)$ 
and around $\Phi= 70^{\rm o}-95^{\rm o}\, (\tan\beta=10, \kappa=2.0)$ 
where $H_1$ is a dominantly CP--odd state.

\vskip -1.cm

\begin{center}
\begin{figure}[htb]
  \vspace*{-1.0cm}
  \hspace*{3.0cm} 
  \epsfxsize=12cm \epsfbox{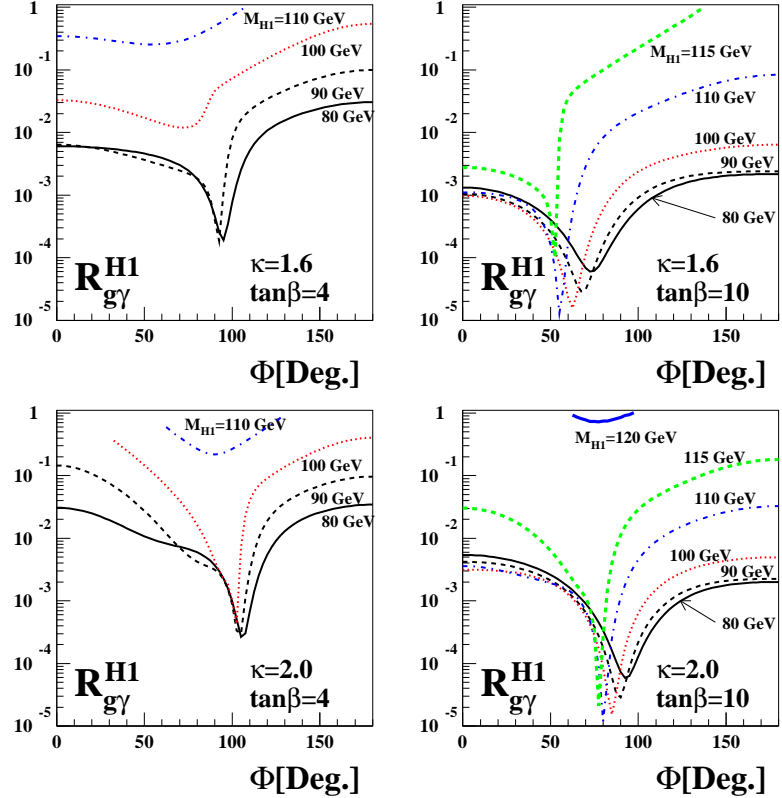}
  \vspace*{-.3cm}
\caption{\it The ratio $R_{g\gamma}^{H_1}$ as a function of the 
         phase $\Phi$ for $m_{H_1}=80$ GeV (solid
         line), 90 GeV (dashed line), 100 GeV (dotted line), 110 GeV 
	 (dash--dotted line), 115 GeV (thick dashed line), and 120 GeV 
	 (thick solid line). The parameter set (\ref{eq:PARA}) is
	 taken with $\kappa=1.6$ (upper) and $2.0$ (lower) and with
         $\tan\beta=4$ (left) and 10 (right). }
\label{r1}
\end{figure}
\end{center}

Since gluon fusion is the main production mode for $H_1$
and the decay $H_1 \rightarrow \gamma \gamma$ is the major signal mode for
$H_1$ at the LHC for $m_{H_1}\leq 130$, it is crucial to investigate 
the observability of the lightest MSSM Higgs boson with explicit CP 
violation through this channel at the LHC. For this purpose, we consider the 
ratio of the signal cross sections:
%
\begin{eqnarray}
\label{eq:ratio}
R_{g\gamma}^{H_i}\equiv\frac{\hspace {-2 cm}\left[\,\hat{\sigma}_{\rm LO}
       (gg\rightarrow H_i)\times {\cal B}(H_i\rightarrow 
       \gamma\gamma)\right]_{\rm MSSM}}{\hspace{0 cm}
       \left[\hat{\sigma}_{\rm LO}(gg\rightarrow H_{\rm SM})
       \times {\cal B}(H_{\rm SM}\rightarrow \gamma\gamma)\right]_{\rm SM}
       \left.\vphantom{\frac{\frac{1}{1}}{1}}
       \right|_{m_{H_{_{\rm SM}}} =m_{_{H_i}}}} \,.
\end{eqnarray}
It measures the $H_i$ signal cross section of the MSSM as compared to the SM
Higgs boson signal cross section at the same mass.
Figure~\ref{r1} shows the ratio as a function of the phase $\Phi$ for 
$m_{H_1}=80-120$ GeV, $\kappa=1.6$ and $2.0$, and  $\tan\beta=4$ and 10, 
as in the previous figures.

\vskip -0.5cm 
The ratio $R_{g\gamma}^{H_1}$ can be highly suppressed
for a wide range of $\Phi$, which is mainly due to the significant decrease of 
${\cal B}(H_1\rightarrow \gamma\gamma)$ caused by
the suppressed coupling of $H_1$ to $W^+W^-$ for non--vanishing $\Phi$.
The ratio is strongly suppressed for a wide range of $\Phi$ around 
$90^{\rm o}$, although otherwise it is not suppressed and can be as large as 
the unity for some cases. For the cases studied in this letter, we find that
the ratio is almost always less than unity.

\begin{center}
\begin{figure}[htb]
  \vspace*{-.5cm}
  \hspace*{3.0cm} 
  \epsfxsize=12cm \epsfbox{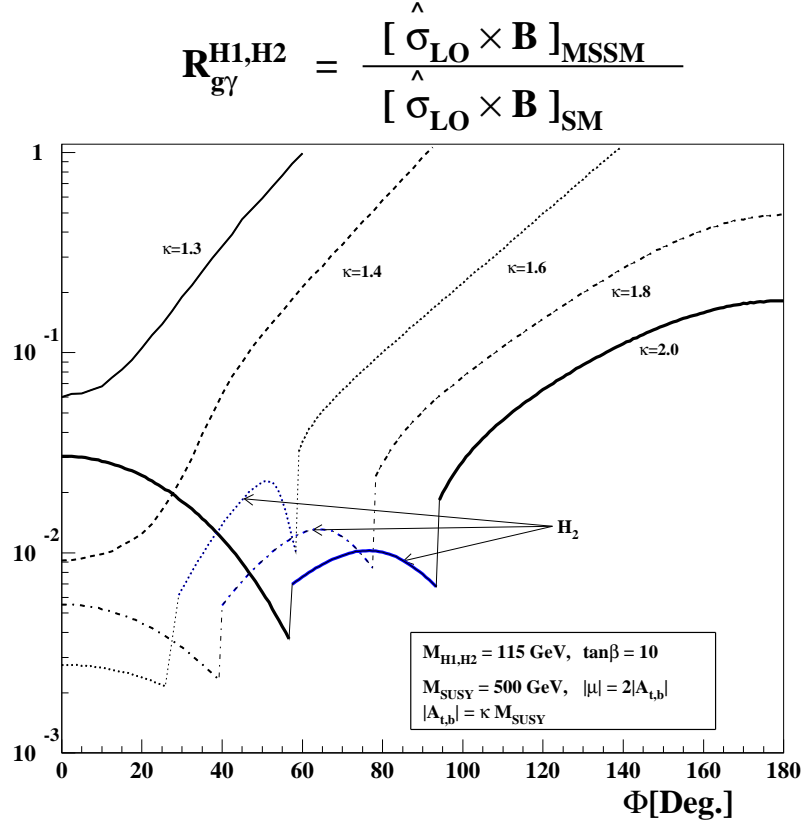}
  \vspace*{-.3cm}
\caption{\it The ratio $R_{g\gamma}^{H_1,H_2}$ as a function of the 
         phase $\Phi$ for $\kappa=1.3$ (solid line),
         $\kappa=1.4$ (dashed line), $\kappa=1.6$ (dotted line),
         $\kappa=1.8$ (dash--dotted line), and $\kappa=2.0$ (thick solid line).
         The mass of the lightest or the second lightest Higgs boson is 
	 fixed with $m_{H_1,H_2}=115$ GeV.  The parameter set (\ref{eq:PARA}) 
         with $\tan\beta=10$ is taken. }
\label{lep2}
\end{figure}
\end{center}

On the other hand, the case for the heavy neutral Higgs bosons is converse
to that for the lightest Higgs boson; the sum rule
for the couplings $ g_{H_iVV}=c_\beta\,O_{2\,i}
+s_\beta\,O_{3\,i}$ of the neutral Higgs bosons to a gauge boson
pair $V$ ($=W^\pm, Z$) ,
\begin{eqnarray}
\label{sumrule}
\sum_{i=1}^{3}g^2_{H_iVV}=1\,, 
\label{eq:orthogonal}
\end{eqnarray}
implies that if $g_{H_1VV}$ is suppressed, either $g_{H_2VV}$ or $g_{H_3VV}$ for
$H_{2,3}$ should be enhanced. Based on the sum rule 
(\ref{eq:orthogonal}), it has been argued \cite{CEPW2} 
that the tantalizing hints for the Higgs boson(s) with its mass
around 115 GeV at the LEP experiments \cite{LEPHiggs} might be due to the 
intermediate or the heaviest Higgs boson instead of the lightest Higgs 
boson which can be CP--odd. In this light, it is worthwhile 
to simultaneously consider the ratios $R_{g\gamma}^{H_{2,3}}$ with 
$m_{H_{2,3}}$ around 115 GeV as well as $R_{g\gamma}^{H_1}$. Let us examine 
the dependence of $R_{g\gamma}^{H_i}$ on the parameter 
$\kappa=|A_{t,b}|/M_{\rm SUSY}$ in the parameter set (\ref{eq:PARA})
for $m_{H_i}=115$ GeV and $\tan\beta=10$.
We find that it is impossible for $H_3$ to be the 115 GeV Higgs boson for the
LEP2 excess events in our parameter set (\ref{eq:PARA}) with 
$\kappa\geq 1.2$. Although, the coupling $g_{H_3VV}$ may dominate the other 
couplings for $\kappa<1.2$, 
we do not consider this case in the present work because we do not find
significant CP--violating mixing for $\kappa<1.2$.
We find that the coupling $g_{H_2VV}$ is 
larger than the coupling $g_{H_1VV}$ in the following parameter space:
\begin{eqnarray}
&& 30^{\rm o}\leq \Phi \leq 60^{\rm o}\hspace{1.7 cm}
   {\rm for}\hspace{0.5 cm} \kappa=1.6\,, \nonumber \\
&& 40^{\rm o}\leq \Phi \leq 80^{\rm o}\hspace{1.7 cm}
   {\rm for}\hspace{0.5 cm} \kappa=1.8\,, \nonumber \\
&& 60^{\rm o}\leq \Phi \leq 100^{\rm o}\hspace{1.5 cm}
   {\rm for}\hspace{0.5 cm} \kappa=2.0\,.
\label{eq:H2dominate}
\end{eqnarray}
In this regard, we present the ratio $R_{g\gamma}^{H_2}$ with $m_{H_2}=115$ GeV 
instead of $R_{g\gamma}^{H_1}$ in the parameter space (\ref{eq:H2dominate}).
Figure~\ref{lep2} shows the ratio $R_{g\gamma}^{H_1,H_2}$ with
$m_{H_1,H_2}=115$ GeV as a function of the phase $\Phi$ for $\kappa=1.3$ 
(solid line), $\kappa=1.4$ (dashed line), $\kappa=1.6$ (dotted line),
$\kappa=1.8$ (dash--dotted line), and $\kappa=2.0$ (thick solid line).
Note that the ratio $R_{g\gamma}^{H_1}$ with $m_{H_1}=115$ GeV
depends strongly on the phase $\Phi$ independently of the value of $\kappa$.
The ratio $R_{g\gamma}^{H_1,H_2}$ becomes smaller for larger $\kappa$.
The ratio $R_{g\gamma}^{H_2}$ with $m_{H_2}=115$ GeV is about $0.01$ in the 
range (\ref{eq:H2dominate}).

It is known \cite{Orsay} that for the integrated luminosity of
100 ${\rm fb}^{-1}$ the signal significance for the discovery of
the SM Higgs boson with $m_{H_{\rm SM}}\leq 130$ GeV through  
the process $gg\rightarrow H_{\rm SM}\rightarrow \gamma\gamma$ 
is less than 10 per experiment at the LHC. It means that
the $5\sigma$--level discovery of the lightest MSSM Higgs boson 
may not be possible at the LHC through this channel if
the ratio $R_{g\gamma}^{H_1}$ is significantly less than a quarter.
As shown in Figs.~\ref{r1} and \ref{lep2} 
the lightest Higgs boson of the MSSM can escape detection
if $\kappa\geq 1.7$.
On the other hand, it may be possible to
discover the lightest Higgs boson at the LHC if $\kappa$ is less than
1.7 and the phase $\Phi$ is sufficiently large.

Let us elaborate on the ratios $R_{g\gamma}^{H_2}$ a little more.  
For a fixed $m_{H_1}$ there should be an anti--correlation
between $R_{g\gamma}^{H_1}$ and $R_{g\gamma}^{H_2}$ due to the sum rule
(\ref{sumrule}) for $g^2_{H_1VV}+g^2_{H_2VV}
\approx 1$; if the coupling $g_{H_1WW}$ is suppressed, the coupling 
$g_{H_2 WW}$ is enhanced and the mass difference $m_{H_2}-m_{H_1}$ is also
reduced. Nevertheless, the ratio $R_{g\gamma}^{H_2}$ for $m_{H_2}\leq 150$ 
GeV is always less than $0.1$ in spite of the anti--correlation
for the parameter space under consideration. It is therefore possible that LHC
discovers neither $H_1$ nor $H_2$ at the LHC for a wide range of
$\Phi$. On the other hand, all the cases shown in Fig.~\ref{lep2}
give either $g^2_{H_1ZZ}$ or $g^2_{H_2ZZ}$ greater than 0.5 so that one of
their production cross sections at LEP2 is not suppressed significantly.
If we require that ${\rm max}\{g^2_{H_iZZ}\}>0.9$, then we find
that the minimum of the ratio is around $1/40$.

%
\vskip -0.5cm 

To summarize, we have investigated the observability of the lightest Higgs 
boson at the LHC by studying its production through gluon fusion and its decay 
into two photons in the MSSM where the tree-level CP invariance in the
Higgs sector is explicitly broken by the loop effects of third--generation 
squarks with CP--violating phases. 
We find that both the production cross section and the decay
branching fraction can be strongly suppressed for non--trivial phase $\Phi$ 
and for large $\kappa$, while the maximal signal cross section
is always for $\Phi=180^{\rm o}$. {\it Consequently, it is possible
that the lightest MSSM Higgs boson escapes detection through the gluon fusion
and its decay into two photons at the LHC if the CP--violating mixing is
significant.} It is therefore important to study seriously the vector--boson
fusion signal at the LHC \cite{ZEPP}.

\subsubsection*{Acknowledgments}

The work of SYC was supported by the Korea Science and Engineering Foundation
through the KOSEF--DFG exchange program (Grant No. 20015--111--02--2) and 
the work of JSL was supported by the Japan Society for the Promotion of 
Science (JSPS).

\bigskip
\bigskip 

\end{document}